\documentstyle[12pt]{article}
\begin{document}

\begin{center}
{\large
{\bf
    Superconducting Fluctuations  and the Pseudogap  in  the
Slightly-overdoped High-$T_{c}$ Superconductor
TlSr$_{2}$CaCu$_{2}$O$_{6.8}$: High Magnetic Field NMR Studies}}

\vspace{3mm}

Guo-qing Zheng $^{1}$, H.Ozaki $^{1}$, W.G. Clark $^{2}$, Y. Kitaoka
$^{1}$, P. Kuhns $^{3}$, A.P.Reyes$^{3}$,   W.G. Moulton $^{3}$, T.
Kondo$^{4}$, Y.Shimakawa$^{4}$ and Y.Kubo$^{4}$
\vspace{2mm} \\
$^{1}$ {\em  Department of Physical Science,  Osaka University, Toyonaka, Osaka 560-8531, Japan.}

$^{2}$ {\em Department of Physics and Astronomy, University of California
at Los Angeles,
CA 90095-1547.}

$^{3}$ {\em National High Magnetic Field Laboratory, Tallahassee, FL
32310.}

$^{4}$ {\em Basic Research Laboratory, NEC Corp. Tsukuba 305-8501,  Japan.}

\vspace{2mm}

(received   Jan. 19, 2000 )

\end{center}

\baselineskip 8mm

  From  measurements of the $^{63}$Cu Knight shift ($K$) and the nuclear
spin-lattice relaxation rate ($1/T_{1}$) under  magnetic fields  from zero
up to 28 T in the slightly overdoped superconductor
TlSr$_{2}$CaCu$_{2}$O$_{6.8}$ ($T_{c}$=68 K), we find that  the pseudogap
behavior, {\em i.e.}, the reductions of $1/T_{1}T$ and $K$ above $T_{c}$
from the values expected from the normal state  at high $T$, is strongly
field dependent and follows  a scaling relation. We show that this scaling
is consistent with the effects of  the    Cooper pair density fluctuations.
The present finding contrasts sharply with the pseudogap property reported
previously  in the underdoped regime   where no field effect was seen   up
to  23.2 T. The implications are discussed.

\vspace{2mm} 


\newpage

The   high transition temperature ($T_{c}$) superconductors have
attracted enomous attention because of their high 
$T_{c}$ and their anomalous normal-state properties. It is believed that
the unusual normal-state properties are due to  strong electron-electron
correlation effects. On the other hand,  it is also pointed out that  some
effects associated with the high value of  $T_{c}$, such as superconducting
fluctuations (SF),  may also complicate the normal-state properties [1].
Among various unsettled issues of the normal state, the 
so-called pseudogap (PG),  which is a phenomenon of spectral weight suppression, has attracted much attention
in recent years. Although the PG is observed  in  most
underdoped materials and possibly also in the overdoped regime [2], its detailed properties remain to be characterized.
Measurements under strong magnetic fields may help  to   discriminate
between different mechanisms that are responsible for the PG [3-6].

In this Letter, we report the temperature ($T$) and magnetic field ($H$)
dependence of the normal-state properties probed by the $^{63}$Cu Knight shift
($K$) and the nuclear spin-lattice relaxation rate ($1/T_{1}$) measurements in the
slightly overdoped superconductor  TlSr$_{2}$CaCu$_{2}$O$_{6.8}$, at both
zero magnetic  field (NQR) and high  fields up to 28 T. It was found that
the  PG behavior is seen but it depends strongly on
$H$. We
further find that  the PG follows
a $T$- and $H$- scaled  relation, which is shown to be consistent with the Cooper pair density fluctuations.    The
present finding contrasts sharply with the PG property   in the
underdoped regime   where no field effect was seen   up to  23.2 T [3].
 Implications of these findings are discussed.

TlSr$_{2}$CaCu$_{2}$O$_{7-\delta}$ consists of two identical CuO$_{2}$ planes in the unit cell.
  The doping level
is controled by changing the oxygen content  by annealing [7]. The as-grown
sample   is
non-superconducting with  $\delta=0.12$. Superconductivity is obtained and $T_{c}$
is increased monotonically  to 70 K  [7, 8] when  $\delta$ is increased,
thereby reducing the carrier concentration. The electrical
resistivity follows a simple power law $\rho=\rho_{0}+aT^{n}$. The exponent $n$ changes gradually from $\sim$1.3
for the highest-$T_{c}$ sample to 1.7 for the as-grown sample [7].  Even  the
sample with the highest $T_{c}$ is suggested to be still in the slightly
overdoped regime [8]. The sample used in this study has a zero-field
critical transition temperature $T_{c0}$ of 68 K with $\delta\sim$0.20 and  $n$=1.3 [7].
All NMR measurements
were done on the central transition (the -1/2 $\leftrightarrow$1/2
transition) in a c-axis aligned powder sample [8]. $1/T_{1}$ was obtained from the recovery of the magnetization
($M(t)$) following a single saturation pulse and a good  fitting    to
$\frac{M(\infty)-M(t)}{M(\infty)}=0.9\exp{(-6t/T_{1})}+0.1\exp{(-t/T_{1})}$
[9].  The transition temperature
    for H$\parallel$ c-axis, $T_{cH}$, was determined from the ac
susceptibility  by measuring the inductance of the NMR coil and  was    58 K,  51 K, 40 K and 37 K for $H$= 7 T, 15.6 T, 23 T
and 28 T, respectively,  as indicated by the arrows in Fig. 1.  Application of   the Werthamer-Helfand-Hoenberg
theory  [10] indicates that a field of 43 T should destroy the
superconductivity completely. This relatively small critical field, $H_{c2}(0)$, is another manifestation of the sample being overdoped.

  Figure 2(a) shows $1/T_{1}T$ as a function of $T$ for 0$\leq H\leq$28 T
parallel to the c-axis.   Figure 2(b) shows the $T$ variation of the Knight
shift ( $K_{c}$ ) for various H$\parallel$ c-axis. Figure 3 emphasizes the
data near $T_{cH}$.  The 
arrows, from right to left,  indicate $T_{cH}$ as
$H$ is increased. At $H$=0,  $1/T_{1}T $ increases with decreasing
$T$  down to $T^{*}$=85 K.  The curve in Fig. 2(a) is  a fitting of the data above $T$=90
K to the relation of
$1/T_{1}T =  \frac{C}{T+\theta} $,
with C=4.7 msec$^{-1}$ and $\theta$=235 K. This Curie-Weiss (CW) relation
of $1/T_{1}T$ was reported in many other high-$T_{c}$ cuprates such as
La$_{2-x}$Sr$_{x}$CuO$_{4}$ [11] and is explained theoretically as caused 
by antiferromagnetic (AF) spin fluctuations [12, 13].
The deviation of $1/T_{1}T$ from the CW relation below $\sim T^{*}$ has
been widely attributed to the loss of low-energy spectral weight, {\em
i.e.},  the opening of a PG.
Applying a magnetic field shifts $T^{*}$  to lower $T$;  $1/T_{1}T$ is
strongly field dependent below $T\sim T^{*}$=85 K.

 $K_{c}$ is the sum of an $H$-and $T$-independent orbital part
$K_{orb}$ and the spin part $K_{s}$, which is  proportional to the uniform spin
susceptibility $\chi_{s}$. In Y- and La-based cuprates $K_{s}$ of $^{63}$Cu for $H\parallel$ c-axis is negligibly small due to accidential cancellation of the hyperfine field, which prevents investigation of the $T$- and $H$-dependence of $\chi_{s}$ at this field alignment.  $K_{s}$ is finite in the present case,  likely due to larger transferred-hyperfine field coming from the nearest Cu [8].   At high $T$, as seen in Fig. 2(b), $K_{c}$ increases slighly with decreasing $T$,
but remains constant at 1.42\%  for  85K$\leq  T \leq$150 K. Such a
$T$-independent $K$  is a common feature of the optimally-doped [14] or
overdoped materials [15]. As seen in Fig. 3(b), $K_{c}$ starts to decrease
at temperatures  above $T_{cH}$, with no apparent singularity at $T_{cH}$.
Like $1/T_{1}T$, the temperature at which $K_{c}$ starts to deviate from a
constant value depends on $H$.

The key experimental results are: (1) The PG temperature at which
$K_{c}$ starts to deviate from a constant value and $1/T_{1}T$ deviates
from that expected by the  relation of $1/T_{1}T = \frac{C}{T+\theta}$, is
lowered progressively by $H$. (2) The reduction of  $K_{c}$ and
$1/T_{1}T$ at $T_{cH}$  become larger  as $H$ is increased. Thus, at a
first glance, it appears that the  PG behavior becomes prominent at
high fields, although it starts at lower $T$.

We find that the $T$- and $H$-dependence of these
reductions follow a scaling relation.
In Fig. 4(a), we show  the normalized reduction of $1/T_{1}T$,
$ \delta (T_{1}T)^{-1}=\frac{(T_{1}T)_{N}^{-1}-(T_{1}T)^{-1}_{obs}}
{(T_{1}T)_{N}^{-1}}$  , divided by  $\sqrt{T_{c0}-T_{cH}}$,  as a function
of the reduced temperature difference, $\frac{T-T_{cH}}{T_{c0}-T_{cH}}$.
Here $(T_{1}T)_{N}^{-1}= \frac{4.7}{T+235}$ in  msec$^{-1}$K$^{-1}$. It is
seen  that the data for four fields  are collapsed onto  a universal curve
down to well below $T_{cH}$.
In Fig. 4(b), the reduction of $K$, $\delta K=1.42\%-K_{c}$
divided by $\sqrt{T_{c0}-T_{cH}}$ is plotted against the   same normalized
temperature difference.
A scaling relation is also evident. In both cases, the 
scaling has  the same dependence on $\frac{T-T_{cH}}{T_{c0}-T_{cH}}$.

We now argue  that these scaling relations are consistent with the effects
of  the fluctuating Cooper pair density. The spin Knight shift is written
as $K_{s}\propto \chi_{s}$;
$1/T_{1}T$ can be written as $1/T_{1}T \propto \sum_{q}\frac{Im \chi(q,\omega)}{\omega}|_{\omega\rightarrow 0}\simeq\chi_{s}\sum_{q\sim Q}\frac{\xi_{M}^{4}}{(1+q^{2}\xi_{M}^{2})^2}$,  where $Im$ means imarginary part, 
$\xi_{M}$ is the magnetic correlation length   and $Q$ is the AF wavevector [12,13].
  $\chi_{s}$ is proportional to the density of states (DOS) or the number of
the normal-state electrons. SF modify 
$\chi_{s}$, and also $\xi_{M}$ in general. 
The effects of SF on various physical quantities
have been extensively studied in the past in conventional superconductors
[16]  and  recently also in high-$T_{c}$ superconductors [17].
Three processes are known: (1) the  Aslamazov-Larkin (AL) term, which is a
direct effect of the  Cooper pairs  formed above mean-field $T_{c}$ [18],
(2) the Maki-Thompson (MT) term, which is due to the coherent scattering of
two counterparts of a Cooper pair  on the same elastic impurities [19], and 
(3) the  DOS term due to the reduction of one-electron DOS
because a part of electrons form Cooper pairs [20, 21].
  The AL contribution to the Knight shift and the relaxation is negligible
because of  singlet  electron pairing.   The MT term is sensitive to the
pairing symmetry but the DOS term is not. When the  Cooper pair is of
$d$-wave symmetry,  which is believed to be realized in the high-$T_{c}$
superconductors, the MT term is much smaller than the DOS term [22, 23].
Under these circumstances,
the predominant effect of the SF on $K$
and $1/T_{1}$ is to reduce the contributions due to the  DOS.

This reduction can be modeled by calculating the corresponding number of the fluctuating 
Cooper pairs, $N_{c.p.}$ which gives rise to the reduction of the DOS. The static  fluctuation of the Cooper pair 
density above $T_{cH}$ can be estimated from
the  Ginsburg-Landau (GL) theory. The GL free energy density relative to
the normal state  is
$f  =  \alpha  |\psi|^{2}+\frac{1}{2m^{*}} |(\frac{\hbar}{i}\nabla-\frac{e^{*}}{c}\vec{A})\psi|^{2}+\frac{\beta}{2}|\psi|^{4}$,
where $\psi$ is the order parameter, $\alpha$ and $\beta$ are constants,
and $m^{*}$  and $e^{*}$ are  the mass  and charge of the Cooper pair,
respectively.
The probability for each $\psi (r)$ is proportional to $exp(-f/k_{B}T)$.
Consider $T$ far enough above $T_{c}$ that the $|\psi|^{4}$ term can be
neglected. Suppose $H\parallel z $ and expand $\psi$ in terms of the wave
function of the Landau orbit $\varphi_{n,k_{z}}$,
$\psi (r) =  \sum C_{n,k_{z}} \varphi_{n,k_{z}}$.  It can be shown  [16] that
  $f= \sum \frac{\hbar^{2}}{ 2m^{*} } [(k_{z}^{2} + \frac{1}{\xi^{2}}+
(n+\frac{1}{2})\frac{4\pi H}{\phi_{0}}] |C_{n,k_{z}}|^{2}$,
where $\phi_{0}$ is the flux quantum  and $\xi\equiv
\sqrt{\frac{\hbar^{2}}{2m^{*}\alpha}}\equiv\xi_{0}/\sqrt{\varepsilon}$ is the GL
coherence length, with  $\varepsilon=log(T/T_{c0})\approx
\frac{T-T_{c0}}{T_{c0}}$. Therefore,  the averaged fluctuation is
$<|\psi_{k_{z},n} |^{2}>=  \frac{k_{B}T}{
\frac{\hbar^{2}}{2m^{*}}[k_{z}^{2} +\frac{1}{\xi^{2}}+
(n+\frac{1}{2})\frac{4\pi H}{\phi_{0}}] }$, where $n$ labels  the Landau
level.
By introducing $\varepsilon_{H}=\frac{T-T_{cH}}{T_{c0}}$, where $T_{cH}$ is
the mean-field transition temperature at field $H$, and
$\tilde{H}=H/H_{c2}(0)$, the averaged fluctuation becomes,
\begin {eqnarray}
<|\psi_{k_{z},n} |^{2}> & = & \frac{k_{B}T}{
\frac{\hbar^{2}}{2m^{*}\xi_{0}^{2}}[\varepsilon_{H} +\xi_{0}^{2}k_{z}^{2}+
2n\tilde{H}]} .
\end{eqnarray}

The factor of $T$ shown in eq. (1)  is cancelled out when one includes the 
effect of dynamic fluctuations (non-zero frequencies). Dynamical fluctuations  suppress the order parameter modulus. This suppression is larger for higher $T$ [16,20,23]. Calculating these fluctuations is rather 
elaborate and has not been derived analytically. However, by using the 
Matsubara Green's function and numerical calculation, Heym [20] found that the 
inclusion of dynamical fluctuations is
equivalent to multiplying the
static fluctuation term by the factor $1/T$ in the temperature range of 
1.05$\leq T/T_{c} \leq 1.6$,
which corresponds to 0.1$\leq \frac{T-T_{cH}}{T_{c0}-T_{cH}} \leq$ 1 in the
present case. On the basis of this argument, we drop the factor $T$ in eq. 
(1) from further consideration.
Then, $N_{c.p.}$ is,
\begin {eqnarray}
N_{c.p.} = \sum_{k}<|\psi_{k} |^{2}> = \int \frac{dk_{z}}{2\pi}
\frac{H}{\phi_{0}}\sum_{n} <|\psi_{k_{z},n} |^{2}> \nonumber \\
 \propto  \sum_{n} \frac{\tilde{H}}{\sqrt{ \varepsilon_{H}+2n\tilde{H}} }  .
\end{eqnarray}

  By dividing the two sides of  eq. (2) by $\sqrt{\tilde{H}}$, one obtains a
  scaling relation between $N_{c.p}/\sqrt{\tilde{H}}$ and
$\varepsilon_{H}/\tilde{H}$. By noting that $H_{c2}(T)$ is linear near
$T_{c}$ so that $\tilde{H}$ can be written as $\tilde{H}
=\frac{1}{0.69}\frac{T_{c0}-T_{cH}}{T_{c0}}$ [10], we obtain,
\begin {eqnarray}
\frac{N_{c.p.}}{\sqrt{T_{c0}-T_{cH}} }& \propto &
\sum_{n}\frac{1}{\sqrt{\frac{T-T_{cH}}{T_{c0}-T_{cH}}+\frac{2n}{0.69}}} .
\end{eqnarray}

Eq.(3) reproduces the scaling found
experimentally. The solid curves in Fig. 4 are fits to   the  right
hand of eq. (3) by taking
$n_{max}=1/\tilde{H}=5$ following ref. [24].  Taking larger $n_{max}$ is found to 
result only in a shift of the fitted region to lower temperature, as found by
Eschrig {\em et.al.} [23].  We have limited the fitting  to the vicinity of
$T_{cH}$ where the GL theory is valid. In the range of 0.1$\leq
\frac{T-T_{cH}}{T_{c0}-T_{cH}} \leq$ 0.35, the fitting is reasonably good.
The deviation from the  GL theory near (and below) $T_{cH}$ is not
surprising since  the fluctuations there ({\em i.e.}, in the
critical region) are strong   that
the $|\psi|^{4}$ term can not be neglected.

Thus, the $H$ and $T$ dependences of  $1/T_{1}T$ and the Knight shift can
be understood as due to   Cooper pair density fluctuations of both static
and dynamic origins.  The reduction of both $K$ and $1/T_{1}T$ are
proportional to $N_{c.p.}$,   as expected by the theory to the first order
of approximation ({\em i.e.}, neglecting the change in $\xi_{M}$).
 The magnetic
field-enhanced fluctuation near $T_{cH}$ is  attributed to  the increase of the
density of the fluctuating  pairs due to the degenerated  Landau  level.

 To our knowledge, this is the first report of the observed scaling relation 
for the NMR quantities. It is observed over a wide
$T$ range that extends well below $T_{cH}$ and up to as high as 
2$T_{cH}$.
Although the above simplified argument gives an intuitive account for the
scaling above (and near) $T_{cH}$, theories which can describe the whole
$T$ range including the critical and the high-$T$
regions are needed.

Finally we discuss  the implications of the present finding. First, the
present experiment  warns that  care should be taken when making
quantitative arguments about the  PG phenomenon under magnetic
field.
As seen in our experiment, applying a magetic field gives a result that is
consistent with the quantization of the orbital motion of the electrons
which in turn lowers $T^{*}$ and makes the  PG
appear more pronounced than at zero field. Secondly,
the PG behavior    in the overdoped regime contrasts sharply
with what is seen in the underdoped material YBa$_{2}$Cu$_{4}$O$_{8}$
reported previously, where no field dependence was found up to 23.2 T, even
though the field reduces $T_{c}$ by 20 K (26\% of $T_{c0}$) [3]. There exists a wide variety of interpretations for  the PG [25] and its $H$-independence [26].  The present findings imply that any electron correlation-driven pseudogap, if realized in the underdoped regime, would  terminate at some doping level  [27],  before entering the overdoped regime. 

In summary, we have found  in the slightly overdoped superconductor
TlSr$_{2}$CaCu$_{2}$O$_{6.8}$ that the reduction of the Knight shift and
$1/T_{1}T$ above $T_{c}$, {\em i.e.}, the pseudogap behavior,  is strongly
field dependent and follows a simple scaling relation.  Based on the GL
theory we argued that this scaling is consistent with  the effects  of the  Cooper pair fluctuations above
mean-field $T_{cH}$. The present results contrast sharply 
with the  pseudogap  property in the underdoped
regime,  which is not affected by a field up to 23.2 T [3]. These results  imply that an electron correlations-driven pseudogap   would terminate at some doping level before entering the
overdoped regime.

   We thank K. Miyake for helpful discussions and comments. Thanks are also
due to S. Fujimoto,  R. A. Klemm,  H. Kohno, O. Narikiyo and  K. Yamada
 for useful discussions.
Partial support by  Grant-in-Aids for
Scientific Research Nos.11640350 (G.-Q.Z), 10044083 and 10CE2004 (Y.K), from
the Ministry of
Education, Science, Sports and Culture, and by NSF grant 
DMR-9705369 (W.G.C) is gratefully acknowleged.    A portion of this work was
performed at National High Magnetic 
Field Laboratory, which is supported by NSF Cooperative Agreement
No. DMR-9527035 and by the State of Florida.

\vspace{3mm}

\begin{flushleft}

 \newpage

{\bf  References:}

\vspace{3 mm}

[1] for recent progress, see {\em Proceedings of the 5th Intl. Conf. on
Materials and Machanisms of Superconductivity and High-$T_{c}$
Superconductors}, eds. Y.-S. He {\em et al}, Physica {\bf C282-287},
(1997).

[2] H. Ding {\em et al},  Nature (London){\bf 382}, 51 (1996).
  A.G.Loeser {\em et al},  Science  {\bf 273}, 325 (1996). CH. Renner {\em et al}, Phys. Rev. Lett. {\bf 80}, 149  (1998). K.  Ishida {\em et al}, Phys. Rev. {\bf B58}, R596 (1998).

[3] G.-q. Zheng {\em et al}, Phys. Rev. {\bf B60}, R9947 (1999).

[4] K.Gorny {\em et al}, Phys. Rev. Lett. {\bf 82}, 177(1999).

[5]  V.F.Mitrovic  {\em et al}, Phys. Rev. Lett. {\bf 82}, 2784  (1999). H.
N. Bachman {\em et al}, Phys. Rev.  {\bf B60}, 7591 (1999).

[6] P. Carretta {\em et al}, Phys. Rev. {\bf B61}, 12420 (2000). {\em ibid} {\bf B54}, R99682 (1996).

 [7] Y. Kubo {\em et al},  Phys. Rev. {\bf B45}, 5553 (1992).

[8] K.Magishi {\em et al}, Phys. Rev. {\bf B54} (1996) 10131. G.-q. Zheng
{\em et al},  Physica {\bf B 186-188} 1012 (1992).

[9] A. Narath, Phys. Rev. {\bf 162}, 320 (1967).

[10]  N.R.Werthamer,  E.Helfand and P.C.Hoenberg,  Phys.Rev.,  {\bf 147},
295 (1966).

[11] Y.Kitaoka {\em et al}, Physica  {\bf C 170}, 189 (1990).

[12] T.Moriya, Y.Takahashi, and K.Ueda, J.Phys.Soc.Jpn. {\bf 59}, 2905 (1990).

 [13]  A.J.Millis, H.Monien and D.Pines, Phys.Rev. {\bf B42}, 167 (1990).

[14] R. E. Walstedt {\em et al}, Phys.Rev. {\bf B45}, 8047 (1992).
M. Horvatic {\em et al}, {\em ibid}, {\bf B48}, 13848 (1993).

[15] G.-q. Zheng {\em et al},  Physica {\bf C208}, 339 (1993).

[16] M.Tinkham, {\em Introduction to Superconductivity}, Second edition,
(McGraw-Hill, Singapore, 1996), Chapt. 8.

[17]  for review, see, A.A.Varlamov {\em et al}, Advances in Physics {\bf
48}, 655 (1999).

[18] L. G. Aslamazov and A. I. Larkin, Phys.  Lett., {\bf 26}, 238 (1968).

[19] K. Maki, Progr. Theoret. Phys. {\bf 39}, 897 (1968).
  R. S. Thompson, Phys. Rev. {\bf B1}, 327 (1970).

[20] J. Heym, J. Low Temp. Phys., {\bf 89}, 869 (1992).

[21] M. Randeria and A.A. Varlamov, Phys. Rev.  {\bf B50}, 10401 (1994).

[22] K. Kuboki and H. Fukuyama, J. Phys. Soc. Jpn. {\bf 58}, 376 (1989).

[23] M. Eschrig, D. Reiner and J.A. Sauls, Phys. Rev. {\bf B59}, 12095 (1999).

[24] V. V. Dorin {\em et al}, Phys. Rev. {\bf B48}, 12951 (1993).

[25] See, {\em e.g.}, references cited in [3] and related articles in [1].

[26] Y. Yanase and K. Yamada, preprint.

[27] The results  on optimally doped
YBa$_{2}$Cu$_{3}$O$_{7-\delta}$ in the literature are controversial.  Recent measurements  (Ref [6] and W. P. Halperin {\em et al}, unpublished) seem to support a pseudogap independent  of field  up to 14.8 T as found by Gorny {\em et al} [4].

\newpage

{\bf
Figure Captions:}

\vspace{2mm}

Fig. 1, AC susceptibility at various fields. The sample was field-cooled.

\vspace{2mm}

Fig. 2, (a) $1/T_{1}T$ of $^{63}$Cu as a function of temperature, $T$, at
various magnetic fields parallel to c-axis. The arrow indicates $T_{c0}$.
The curve is a fitting of the data above $T$=90 K to a Curie-Weiss
relation. (b) $T$-variation of the Knight shift, $K_{c}$, at various
fields. The typical uncertainty of the data points is about $\pm$2\%, which is slightly larger than the size of the symbols.

\vspace{2mm}

Fig. 3, Enlarged part of Fig. 2 around $T_{cH}$. The arrows, from right to
left,  indicate $T_{cH}$ at elevated fields.

\vspace{2mm}

Fig. 4, (a) Normalized reduction of $1/T_{1}T$ from that expected from the
high-$T$ Curie-Weiss  relation, divided by $\sqrt{T_{c0}-T_{cH}}$, in
K$^{-\frac{1}{2}}$, plotted as a function of
$\frac{T-T_{cH}}{T_{c0}-T_{cH}}$.  (b) Reduction of the Knight shift from
the constant value, divided by $\sqrt{T_{c0}-T_{cH}}$, in
10$^{-4}$K$^{-\frac{1}{2}}$,  versus $\frac{T-T_{cH}}{T_{c0}-T_{cH}}$.  The
symbols are the same as those in (a). The curves are   fits to
$\frac{N_{c.p.}}{\sqrt{T_{c0}-T_{cH}}}$ (see  text).

\end{flushleft}

\end{document}